\documentclass{article}
\usepackage{spconf,amsmath,graphicx}
\usepackage{cite}
\usepackage{amsmath,amssymb,amsfonts}
\usepackage{algorithmic}
\usepackage{graphicx}
\usepackage{textcomp}
\usepackage{xcolor}
\usepackage{amsfonts}

\usepackage{tikz}
\usepackage{enumitem}
\usepackage{array, makecell}
\usepackage{multirow}
\usepackage{hhline}
\usepackage{xcolor,colortbl}
\usepackage{cite}
\usepackage{algorithm}
\usepackage{algorithmic}

\title{Spectro Temporal EEG Biomarkers for Binary Emotion Classification}
\name{Upasana Tiwari, Rupayan Chakraborty, Sunil Kumar Kopparapu.}
\address{TCS Research -- INDIA. \\ \small{email:\{tiwari.upasana1, rupayan.chakraborty, sunilkumar.kopparapu\}@tcs.com}}

\begin{document}
%\ninept
%
\maketitle
\begin{abstract}
	Electroencephalogram (EEG) is one of the most reliable physiological signal for emotion detection. Being non-stationary in nature, EEGs are better analysed by spectro temporal representations. Standard features like Discrete Wavelet Transformation (DWT) can represent temporal changes in spectral dynamics of an EEG, but is insufficient to extract information other way around, i.e. spectral changes in temporal dynamics. On the other hand, Empirical mode decomposition (EMD) based features can be useful to bridge the above mentioned gap. Towards this direction, we extract two novel features on top of EMD, namely, (a) marginal hilbert spectrum (MHS) and (b) Holo-Hilbert spectral analysis (HHSA) based on EMD, to better represent emotions in 2D arousal-valence (A-V) space. The usefulness of these features for EEG emotion classification is investigated through extensive experiments using state-of-the-art classifiers. In addition, experiments conducted on DEAP dataset for binary emotion classification in both A-V space, reveal the efficacy of the proposed features over the standard set of temporal and spectral features.
%	EEG signal being non-stationary are better analysed by the tempo-spectral representations. Discrete Wavelet Transformation (DWT) has been one of most popular technique that provides the temporal changes in spectral dynamics of a signal but is insufficient to provide the other way around. Empirical mode decomposition (EMD) being able to estimate the instantaneous frequency over time bridge this gap. In this work, we explore the richness of marginal hilbert spectrum (MHS) and Holo-Hilbert spectral analysis (HHSA) based on EMD, to discriminate the EEG emotion in 2-D arousal-valence (A-V) space.
%	Furthermore, we present the multi-domain feature extraction based on DWT, EMD along with several other temporal and spectral features. 
%	The effective features for EEG emotion classification are investigated by an extensive experiments with three state-of-the-art classifier. Working towards more difficult problem, trial-level EEG emotion recognition as the primary objective, we observed that varied set of explored features to be competitive in recognition task. Thus, we fused the posteriors to make use of the complementary information provided by each feature set. Experiments conducted on DEAP dataset to address the binary emotion classification task in A-V space,  shows that the proposed method is more
%efficient than several methods of comparison.
\end{abstract}


\vspace{-0.2cm}
\section{Introduction}
\vspace{-0.2cm}
Human emotion has recently emerged to bridge the gap between humans-robot and human-computer interactions \cite{picard1995affective}. Even though facial expression and speech signals are two conventional modalities to recognize human emotions, both have several limitations in terms of social and cultural dependency of the subjects. Every psychological reactions which stimulates the human emotion, produces a physiological reaction as well. These physiological signal being involuntary and insusceptible to external environment, can assist in better and reliable understanding of subject's underlying responses to any external stimuli. It has been explored through cognitive science and psychological experiments that electrical activity generated by brain plays a vital role in expressed emotions. Hence, Electroencephalogram (EEG) is one of the reliable physiological signal, which is commonly used to recognize emotion
\cite{chanel2006emotion}.
%,chanel2007valence}.

In the literature, EEG-based emotion recognition task have been focused in various areas, such as feature extraction, classification, selecting channels, artifacts filtering etc. \cite{alarcao2017emotions}. A variety of features have been studied from time domain \cite{liu2013real,hjorth1970eeg}
%takahashi2004remarks
, frequency domain \cite{jenke2014feature,rozgic2013robust}
%wang2011eeg
, and time-frequency domain
for which two of the most common techniques used in the literature are Discrete Wavelet Transformation (DWT) \cite{mohammadi2017wavelet,li2018emotion} 
%murugappan2010classification
and Empirical Mode Decomposition (EMD) \cite{ji2019eeg,zhang2016approach}
%pigorini2011time,uzun2012emotion}. 
%
%There are also studies which focus on using different classifiers and channel selection methodologies
.
%In \cite{yoon2013eeg}, authors proposed a probabilistic classifier based on Bayes' theorem and a supervised perceptron convergence learning algorithm with frequency-domain features.
The work in \cite{hatamikia2014emotion} investigated the autoregressive (AR) features by using a Burg’s method, followed by a classification using K-nearest neighbor (KNN). The authors in \cite{atkinson2016improving} explored a wider set of emotion types
by combining mutual information based feature selection methods and kernel classifiers. 
The work in \cite{kumar2016bispectral} conducted the bi-spectral analysis of EEG using a arousal-valence (A-V) emotion model. 
%The authors in \cite{khosrowabadi2010eeg}
%employed a self-organizing map for boundary detection followed by classification using KNN. 
%In addition to crafting effective features and building efficient classifier, there are few works that propose to use privileged information which embodies the private and public characteristics in EEG signal with respect to different subjects \cite{wu2016employing,gao2015emotion}.

%Recent studies have proposed many deep learning methods in EEG emotion recognition. Kwon et al. \cite{kwon2018electroencephalography} made use of time-frequency feature maps of each channel as well
%as other physiological features with convolutional neural network (CNN) to recognize
%emotion states. Li et al. \cite{li2017human} constructed multi-dimensional feature
%images using frequency-domain features which were then fed to a hybrid deep learning model that integrated CNN and recurrent neural network
%(RNN) models. The work in \cite{chao2019emotion} proposed a deep learning framework based on a multi-band feature matrix (MFM) and a capsule network (CapsNet). The authors in \cite{li2021hierarchical} ivestigates hierarchical attention-based temporal convolutional networks that aggregates the emotional content at both the frame and channel level.

Although the work in \cite{mohammadi2017wavelet,li2018emotion,bazgir2018emotion} shows a remarkable segment-level recognition accuracy in EEG-based emotion recognition 
%,vijayan2015eeg}
, but the common drawback in all is that segments from the same trial have been used in both training and testing. 
Consequently, the performance of same systems degrade drastically when tested on segments from unseen trials. 
Hence, in this paper we address trial independant binary classification.
In this work, we explore the effectiveness of data-driven, adaptive, non-linear signal representation, EMD technique. Besides analysing the Intrinsic Mode Functions (IMFs) extracted from EMD at Marginal Hilbert Spectrum (MHS) level, we here explored one of the most recent technique developed to represent the non-linear signal, known as Holo-Hilbert spectral analysis (HHSA). To best of our knowledge, this is the first attempt to use HHSA for recognizing emotion in EEG. 
Further, we extract the multi-domain feature extraction based on DWT, EMD along with several other temporal and spectral features. 
We hypothesise that both DWT and EMD being complementary in terms of their ability to provide the temporal changes in spectral domain and vice-versa result in much more informative representation. 
The effective features for EEG emotion classification are investigated by an extensive experiments with three state-of-the-art classifier, namely, Support vector Machine (SVM), Random Forest (RF) and KNN. To utilise the complementary information provided by each feature set, we fuse the posteriors of models trained with each of them, which improved the recognition accuracy. Our proposed approach with RF as a classifier achieved weighted F1-score and accuracy of  67.2\% and 68.8\% on valence and 66.6\% and 67.5\% on arousal, respectively.

%The rest of the paper is organized as follows, in section \ref{features}, we explain in detail the each set of explored EEG biomarkers for Emotion Recognition. The experimental details, and results are presented in section \ref{exp}, followed by conclusion in section \ref{conclusion}.

\vspace{-0.2cm}
\section{EEG Biomarkers for Emotion Recognition}
\label{features}
\vspace{-0.2cm}
%In this paper, we study the EMD and DWT based features to fuse the complementary tempo-spectral information provided by both techniques. 
%In addition to that, we also use several time-domain (temporal) and frequency-domain (spectral) features. 

In this section, we discuss the biomarkers explored for EEG emotion classification.

\vspace{-0.2cm}

\subsection{Spectro Temporal Features}
\vspace{-0.1cm}
%\label{tempo-spec}
\subsubsection{\textbf{Discrete Wavelet Transformation (DWT)}}
\vspace{-0.1cm}
\label{DWT}
In DWT, the time-frequency representation is obtained by repeatedly filtering.
% a signal with a pair of filters, (i) high-pass and (ii) low-pass. Single stage of filtering divides the frequency domain into two equal half. 
Precisely, DWT decomposes the signal into two parts at each stage, (i) approximation and (ii) detailed signal. In the subsequent stages, the approximation signal is again decomposed into new approximation coefficient (AC) and the detailed coefficient (DC). This step is repeated which produces set of approximation signal at different detail levels and a final approximation of the signal. Here, we use 'db4' wavelet due to their efficient time-frequency localization properties. Moreover, it's waveform is similar to EEG waveforms. The compact representation provided by extracted DWT coefficients shows the energy distribution of the EEG signal in both time and frequency domain. 
The DCs for signal sampled at 128Hz results into decomposed frequency bands of 4-8Hz, 8-16Hz, 16-32Hz, 32-64Hz, corresponds to standard EEG frequency bands (brainwaves) i.e. Theta ($\theta$), Alpha ($\alpha$), Beta ($\beta$), Gamma ($\gamma$), respectively. The initial DC i.e 64-128Hz corresponds to noise and is dropped.
We extract entropy and energy for each of the frequency bands \cite{mohammadi2017wavelet}. 

\vspace{-0.2cm}
\subsubsection{\textbf{Empirical Mode Decomposition (EMD)}}
\vspace{-0.1cm}
\label{EMD}
EMD is an adaptive algorithm that decomposes a signal into IMFs through an iterative process, known as sifting. Each IMF must satisfy the following two conditions \cite{huang1998empirical}:

\begin{enumerate}
	\item The number of maxima and minima are either equal, or differ at most by one.
	\item The mean value of the local envelope defined by the local maxima and local minima is zero.
\end{enumerate}

%The EMD algorithm keeps decomposing the signal iteratively into series of IMFs, this process is known as sifting. 
At each sifting iteration, highest frequency  component  present in the residue is extracted. The process terminates when residual signal either becomes a constant or signal with a single extrema from which no further IMF can be extracted.
Thus, on the completion of EMD, original signal $x(t)$ can be represented as the sum of $K$ IMFs ($I$) and a residue $res_K(t)$
\begin{equation}
x(t)=\Sigma_{i=1}^{K}I_i(t)+res_K(t)
\end{equation}

The biggest advantage of the IMFs are well-behaved Hilbert transforms that allows the extraction of physically meaningful instantaneous frequencies. 
Unlike Wavelet Transform (WT) and Short Time Fourier Transform (STFT), IMFs have large time-bandwidth products that avoid the limitations of the uncertainty principle. 
Furthermore, EMD being data-driven doesn't require apriori knowledge (time windows, mother wavelets etc.) about the signal that makes it more efficient to analyse the highly non-linear and complex EEG signals. 
%that doesn't have well-defined stationary time-windows 
After IMF computation, each of them are converted into analytic signal %($z_i$) 
with Hilbert–Huang transform (HHT) to extract the instantaneous frequency ($\omega$) and instantaneous amplitude ($a$).
We analyse these IMFs at three levels:
%\begin{itemize}%[leftmargin=*]

\vspace{0.3cm}
\noindent
\textbf{I. IMFs based temporal and spectral features \cite{kaleem2013pathological}}: For each IMFs $I_i$, for $i$=$1$..$K$, we compute set of five features, namely, IMF energy, spread ($SP_i(ITED)$) and deviation ($D_i(ITED)$) of the instantaneous temporal energy density, spread ($SP(\omega_i)$) of the values of $\omega$ and deviation ($D_i(ISED)$) of the instantaneous spectral energy density \cite{kaleem2013pathological}.

\vspace{0.3cm}
\noindent
\textbf{II. MHS \cite{huang1998empirical}}: After applying HHT, once we have the instataneous values ($\omega_i$, $a_i$) for each IMF $I_i$, we compute Hilbert–Huang Spectrum ($H[\omega,t]$) that provides the time-frequency-amplitude distribution and is expressed as Eq. \ref{eq:hht}.
From $H[\omega,t]$, we compute the marginal spectrum ($h[\omega]$)  that gives the total
energy distribution within specified frequency bins over a whole signal length, as expressed in Eq. \ref{eq:mhht} . These frequency bands are defined by minimum frequency ($freq_{min}$), maximum frequency ($freq_{max}$) and number of frequency steps ($n\_bins$). 

\begin{equation}
	\label{eq:hht}
	H[\omega,t]=Re\Sigma_{i=1}^{K}a_i(t)e^{j\int\omega_i[t]dt}
\end{equation}
\begin{equation}
	\label{eq:mhht}
	h[\omega]=\Sigma_{t=1}^{N}H[\omega,t]
\end{equation}
where, $N$ is the length of the time samples.
%exp::	To reduce the dimensional complexity (as the models used requires flattened vector as input), we here extracted 1-D hilbert spectrum (with no time dimension).

\vspace{0.3cm}
\noindent	
\textbf{III. HHSA \cite{huang2016holo}}: This is one of the most recent spectral analysis technique for nonlinear systems that overcome the deficiencies
of conventional spectral analysis and give a full informational representation of nonlinear signal. 
So far, the best possible way to analyse the non-linear signals was to use the time-frequency representations, in which the amplitude (or energy density) variation is still represented in terms of time. However, the non-linear signals often have the presence of intrinsic amplitude and frequency modulations which were left untreated by traditional spectral methods.
HHSA technique uses a nested EMD and HHT approach to identify these intrinsic amplitude and frequency modulations. Holo prefix in HHSA symbolizes a multiple
dimensional representation with both additive and multiplicative capabilities \cite{huang2016holo}. 

The amplitude modulations described by the $a$ are itself oscillatory. Thus, frequency information of these amplitude modulation signal can be described with another EMD. This process is called second-level sift in which $a_i$ (per IMF $I_i$) extracted from first-level sift is further decomposed into another set of IMFs. Thereafter, the second-level frequency stats $a_i^{'}$ and $\Omega_i^{'}$ are computed. Now the holospectrums are computed that describes the distribution of signal power as a function of both frequency of the carrier wave (first-level frequency stats) and the frequency of amplitude modulations (second-level frequency stats).
\vspace{-0.3cm}
%\end{itemize}
%%%%% Symbols and eq are used as it is from \cite{kaleem2013pathological}.
%\vspace{-0.2cm}

\subsection{Temporal and Spectral Features}
\vspace{-0.1cm}
\label{tempo}
In temporal domain, we extracted four types of features \cite{bao2011pyeeg}, namely, (1) Fractal dimension with Higuchi and Petrosian algorithms; (2) Hjorth mobility and complexity parameters; (3) Detrended Fluctuation Analysis (DFA); (4) Hurst Exponent.
%\begin{enumerate} [label=(\alph*)]
%	\vspace{-0.2cm}
%	\item Fractal Dimension: This feature captures information regarding the steepness of the signal and thus represents the shape. Higuchi and Petrosian algorithms are used to extract fractal dimension related features. %%%\cite{higuchi1988approach,petrosian1995kolmogorov}.
%	\vspace{-0.2cm}
%	\item Hjorth Parameters: This temporal features use the statistics like mean and variance of time signal to capture the mobility and complexity of the signal.
%	\vspace{-0.2cm}
%	\item Detrended Fluctuation Analysis (DFA): This feature is used to determine the statistical self-affinity of a signal, thus provide the long range correlation in a non-stationary time-series.
%	\vspace{-0.2cm}
%	\item Hurst Exponent: This is also known as the re-scaled range statistics. This temporal feature is used to measure of the variability of a time series, which captures an assessment of how the apparent variability of a series changes with the length of the time-period being considered.
%\end{enumerate}
%\vspace{-0.2cm}
%\subsection{Spectral Feature}
%\vspace{-0.1cm}
%\label{spec}
The spectral features are extracted from the following frequency bands, namely, $\theta$ (4-8Hz), $\alpha_{low}$ (8-10Hz), $\alpha_{high}$ (10-13 HZ), $\beta$ (13-25Hz) and $\gamma$ (25-40Hz). 
The raw EEG signal is first transformed into frequency domain using fast Fourier transformation (FFT). Thereafter, 3 types of features \cite{bao2011pyeeg, Biosppy} are extracted, namely, (1) Power Spectral Density (PSD); (2) Relative Intensity Ratio (RIR); (3) Spectral Entropy.
%(1) Power Spectral Density (PSD): It computes the average power in each above mentioned frequency bands that indicates the brain-activity; (2) Relative Intensity Ratio (RIR): It is computed as the ratio of spectral power in a frequency band normalized by the total power in all the bands; (3) Spectral Entropy: This feature captures the spectral power distribution in the signal.

\vspace{-0.4cm}
\section{Experiments}
\vspace{-0.2cm}
\label{exp}
\subsection{EEG Database and Data Preprocessing}
\vspace{-0.1cm}
In this work, we use
the Database for Emotion Analysis using Physiological
signals (DEAP) which is having valence-arousal-dominance-liking
emotion dimensions labeled \cite{koelstra2011deap}. 
One-minute long music video clips are used as a audio-visual stimuli to elicit emotions. There are 32 subjects, where each subject was shown 40 clips (trials), and for each trial 7 physiological modalities were recorded. Each trial has physiological recording of 40 channel in which first 32 channels corresponds to EEG. Each subject has self rated the trial between 1 to 9 for each of dimensions.
Here, we have used the preprocessed version \cite{koelstra2011deap} of data in which each trial (signal) data is downsampled from 512Hz to 128Hz and then band pass filtered (4.0-45.0Hz). Originally, each trial has 63sec of data which includes 3sec of pre-trial. In this work, we have taken only 60sec data, dropping the 3sec pre-trial. 
%\subsection{Threshold ($\lambda$) for binary classification}
\begin{table}[t]
	\vspace{-0.6cm}
	\caption{Best performing features (/bands) with reduced dimension ($d$)}% selected for valence-arousal space}
	
	\label{table:FS}
	%	\vspace{-0.2cm}
	\centering
	
	\resizebox{0.5\textwidth}{!}
	{
		\begin{tabular}{|c|c|c|c|c|c|c|}
			\hline
			\multirow{2}{*}{\textbf{Sets}} & \multicolumn{3}{c|}{\textbf{Valence}} & \multicolumn{3}{c|}{\textbf{Arousal}} \\ \cline{2-7} 
			& \textbf{Features} & \textbf{Bands/IMFs} & \textbf{$d$} & \textbf{Features} & \textbf{Bands/IMFs} & \textbf{$d$} \\ \hline
			\textbf{set\_A} & \begin{tabular}[c]{@{}c@{}}Hjorth features, HFD,\\ PFD, Spectral Entropy,\\ PSI, RIR\end{tabular} & $\alpha_{low}$, $\alpha_{high}$, $\beta$, $\gamma$ & 416 & \multicolumn{3}{c|}{\multirow{2}{*}{Same as Valence}} \\ \cline{1-4}
			\textbf{set\_B} & PSDs & $\alpha_{high}$, $\beta$ & 64 & \multicolumn{3}{c|}{} \\ \hline
			\textbf{set\_C} & \begin{tabular}[c]{@{}c@{}}Energy, \\ Entropy\end{tabular} & $\gamma$ & 64 & \begin{tabular}[c]{@{}c@{}}Energy, \\ Entropy\end{tabular} & $\theta$, $\alpha$, $\beta$, $\gamma$ & 256 \\ \hline
			\textbf{set\_D} & \begin{tabular}[c]{@{}c@{}}SP(ITED), \\ D(ISED)\end{tabular} & Top 4 IMFs & 256 & \begin{tabular}[c]{@{}c@{}}IMF Energy,\\ SP($\omega$), \\ D(ISED)\end{tabular} & Top 3 IMFs & 288 \\ \hline
			
		\end{tabular}
		
	}
	\vspace{-0.5cm}
\end{table}

In this paper we address the binary classification task which results by applying threshold ($\lambda$) on the self-assessments provided in \cite{koelstra2011deap}.
The affective label will be set to high if the rating is $>$ 4.5 and low if rating $<=$ 4.5. Thus for each trial, two labels were
generated. V+ (high valence) or V- (low valence); and A+ (high arousal) or A- (low arousal) to get the affective level in valence and arousal space, respectively. 
We addressed the recognition in two dimensions, A-V, as two independent tasks and present them as a two-class problem.

\vspace{-0.2cm}
\subsection{Feature Extraction}
\vspace{-0.1cm}
\label{FE}
In this work, we used five different toolbox for extracting the features discussed in section \ref{features}. We categorize them in six different sets as follows:
\begin{enumerate}
	\item \textbf{set\_A} contains temporal features plus PSI, RIR and Spectral entropy which are extracted using PyEEG \cite{bao2011pyeeg}. All the features computed using PyEEG are at the trial-level without any frame-level computation.
	\vspace{-0.2cm}
	\item \textbf{set\_B} contains PSDs of 5 frequency bands as mentioned in section \ref{tempo} and is computed using BioSPPy Toolbox \cite{Biosppy} with 4sec windows and 2sec overlap.
	\vspace{-0.2cm}
	\item \textbf{set\_C} contains spectro temporal features computed using DWT as discussed in section \ref{DWT} with 4sec windows and 2sec overlap, 
%	thus giving 29 windows for each trials 
	followed by decomposition of each window into 5 levels with PyWavelets tool \cite{Lee2019} and thus, retaining all frequency components as 4 frequency bands (excluding the first detailed component as noise)
%	, namely, $\gamma$, $\beta$, $\alpha$ and $\theta$ 
	as discussed in section \ref{DWT}. Finally, energy and entropy are computed in each band .
	\vspace{-0.2cm}
	\item \textbf{set\_D} contains IMFs based temporal and spectral features. We use PyEMD tool \cite{PyEMD} for decomposing each trials into IMFs. As EMD is data-driven, the number of IMFs varies among trials, we got 10 as the least number of IMFs. So, to maintain the uniformity, 10 IMFs per 1280 trials are considered and then 5 features (as discussed in section \ref{EMD}) per trial are extracted.
	\vspace{-0.2cm}
	\item \textbf{MHS and HHSA} : We use $emd$ \cite{quinn2021emd} python tool to compute the two spectrums, with $freq_{min}$ and $freq_{max}$ as 5Hz and 45Hz to match with standard EEG frequency band limit. For $n\_bins$, we experimented with values ranging from 4 to 128 (higher value gives better frequency resolution). $n\_bins$=64 and $n\_bins$=5 performed the best for MHS and HHSA respectively. It returns a 2D matrix with dimension of $K$*$n\_bin$ for MHS and $n\_bin^{1}*n\_bin^{2}$ for HHSA. Here, $n\_bin^{1}$ and $n\_bin^{2}$ corresponds to first-level and second-level freqency bins as discussed in section \ref{EMD}. 
%	In order to gel with the input format of the classifiers used, both 2-D matrix are converted into flattened vector by concatenating rows of respective matrix.
%	\item \textbf{HHSA} : 
\end{enumerate}

The default frequency bands in PyEEG are modified as BioSPPy's 5 frequency band to make the feature extraction uniform.
All the features in set\_B and set\_C are first computed at the frame-level and later averaged to give the trial-level features.
%Table \ref{table:FD} lists all the features along with their dimensions. 

\vspace{-0.2cm}
\begin{table*}[ht!]
	\vspace{-0.3cm}
	\centering
	\caption{Weighted performance on different sets with feature selection (F1: $F1$ $score$, CA:$Classification$ $Accuracy$)}
	\label{tab:res2}
	\resizebox{0.8\textwidth}{!}
	{
		\begin{tabular}{|l|l|l|l|l|l|l|l|l|l|l|l|l|}
			\hline
			\multicolumn{1}{|c|}{\multirow{3}{*}{\textbf{\begin{tabular}[c]{@{}c@{}}Feature \\ set\end{tabular}}}} & \multicolumn{6}{c|}{\textbf{Valence}} & \multicolumn{6}{c|}{\textbf{Arousal}} \\ \cline{2-13} 
			\multicolumn{1}{|c|}{} & \multicolumn{2}{c|}{\textbf{RF}} & \multicolumn{2}{c|}{\textbf{SVM}} & \multicolumn{2}{c|}{\textbf{KNN}} & \multicolumn{2}{c|}{\textbf{RF}} & \multicolumn{2}{c|}{\textbf{SVM}} & \multicolumn{2}{c|}{\textbf{KNN}} \\ \cline{2-13} 
			\multicolumn{1}{|c|}{} & \textbf{F1} & \textbf{CA} & \textbf{F1} & \textbf{CA} & \textbf{F1} & \textbf{CA} & \textbf{F1} & \textbf{CA} & \textbf{F1} & \textbf{CA} & \textbf{F1} & \textbf{CA} \\ \hline
			\textbf{set\_A} & 65.32 & 65.31 & \textbf{65.80} & \textbf{66.24} & 64.53 & 65.07 & \textbf{65.73} & \textbf{66.56} & 62.81 & 63.41 & 65.33 & 64.93 \\ \hline
			\textbf{set\_B} & \textbf{67.02} & \textbf{67.42} & 61.50 & 61.32 & 61.10 & 61.87 & \textbf{64.84} & \textbf{65.31} & 56.64 & 57.09 & 62.35 & 60.96 \\ \hline
			\textbf{set\_C} & \textbf{65.58} & \textbf{65.93} & 57.22 & 60.15 & 59.18 & 59.76 & \textbf{64.53} & \textbf{64.92} & 60.70 & 61.40 & 61.88 & 60.69 \\ \hline
			\textbf{set\_D} & \textbf{66.24} & \textbf{67.26} & 59.91 & 59.45 & 62.67 & 63.51 & \textbf{64.70} & \textbf{65.85} & 58.28 & 59.07 & 63.12 & 61.95 \\ \hline
			\textbf{MHS} & \textbf{65.52} & \textbf{66.71} & 61.93 & 61.48 & 63.51 & 64.06 & \textbf{65.17} & \textbf{66.33} & 59.77 & 60.37 & 61.09 & 59.85 \\ \hline
			\textbf{HHSA} & \textbf{66.56} & \textbf{67.33} & 58.80 & 58.20 & 62.37 & 62.73 & \textbf{64.91} & \textbf{65.46} & 60.55 & 61.18 & 59.84 & 58.63 \\ \hline
		\end{tabular}
	}
	\vspace{-0.3cm}
\end{table*}

%\begin{table}
%	\vspace{-0.3cm}
%	\caption{Comparison results for emotion recognition on DEAP dataset}%in arousal-valence space}
%	
%	\label{tab:sota}
%	%	\vspace{-0.1cm}
%	\centering
%	
%	\resizebox{0.48\textwidth}{!}
%	{
%		% Please add the following required packages to your document preamble:
%		% \usepackage{multirow}
%		
%			\begin{tabular}{|l|l|l|l|l|}
%				\hline
%				\multirow{2}{*}{\textbf{Features}} & \multirow{2}{*}{\textbf{Classifiers}} & \multirow{2}{*}{\textbf{Evaluation}} & \multicolumn{2}{c|}{\textbf{CA}} \\ \cline{4-5} 
%				&  &  & \textbf{Arousal} & \textbf{Valence} \\ \hline
%				DBN features \cite{li2015eeg} & SVM & 10-fold & 64.2 & 58.4 \\ \hline
%				\begin{tabular}[c]{@{}l@{}}Multiband Feature \\ Matrix \cite{chao2019emotion}\end{tabular} & CapsNet & 10-fold & \textbf{68.3} & 66.7 \\ \hline
%				\begin{tabular}[c]{@{}l@{}}Empirical Wavelet \\ Transform \cite{huang2017eeg}\end{tabular} & SVM & 10-fold & 67.3 & 64.3 \\ \hline
%				\multicolumn{2}{|l|}{\multirow{1}{*}{Our approach (Late fusion)}} & 5-fold & 67.5 & \textbf{68.8} \\ \hline
%			\end{tabular}
%		
%	}
%	\vspace{-0.3cm}
%\end{table}
\begin{table}
	\vspace{-0.3cm}
	\caption{Comparison results for emotion recognition on DEAP dataset}%in arousal-valence space}
	
	\label{tab:sota}
	%	\vspace{-0.1cm}
	\centering
	
	\resizebox{0.48\textwidth}{!}
	{
		% \usepackage{multirow}
		
		% \usepackage{multirow}
		
			\begin{tabular}{|l|l|l|l|}
				\hline
				\multirow{2}{*}{\textbf{Features}} & \multirow{2}{*}{\textbf{Classifiers}} & \multicolumn{2}{c|}{\textbf{CA}} \\ \cline{3-4} 
				&  & \textbf{Valence} & \textbf{Arousal} \\ \hline
				DBN features \cite{li2015eeg} & SVM & 58.4 & 64.2  \\ \hline
				\begin{tabular}[c]{@{}l@{}}Multiband Feature \\ Matrix \cite{chao2019emotion}\end{tabular} & CapsNet & 66.7 & \textbf{68.3}  \\ \hline
				\begin{tabular}[c]{@{}l@{}}Empirical Wavelet \cite{huang2017eeg} \\ Transform\end{tabular} & SVM & 64.3 & 67.3  \\ \hline
				\multicolumn{2}{|l|}{Our approach (with late fusion)} & \textbf{68.8} & \textbf{67.5} \\ \hline
			\end{tabular}

	}
	\vspace{-0.4cm}
\end{table}
\subsection{Features and IMF Selection}
\vspace{-0.1cm}
\label{FS}
%Computing features for all 32 channels and in each frequency bands (as represented in table \ref{features}) results in high dimension feature vector.
Concatenating the features from all 32 channels, results in high dimensional complexity and information redundancy. Thus, we performed 
%most informative 
the feature selection for each of the feature sets discussed in section \ref{FE}. We evaluated each individual features and all possibles subsets of the same. The only subsets performing best or almost similar to the overall set are retained. %and remaining features are dropped.
Also, EMD technique decomposes a signal in a way such that initial IMFs contains the higher oscillation frequencies and it keeps reducing down with further decomposition.
%(as represented in figure \ref{fig-IMF}). 
Hence, we analysed the contribution of each IMFs by considering IMFs in incremental fashion, starting with IMF-1 to IMF-10.
%as represented in Graph \ref{fig-IMF-selection}. 
We found that higher IMFs are more informative and performance starts degrading or saturates as we keep increasing the lower level IMFs. Based on performance, we selected top 4 IMFs for valence and top 3 IMFs for arousal for further experiments.

% \usepackage{multirow}

\vspace{-0.4cm}
\section{Result and analysis}
\label{sec:res}
\vspace{-0.3cm}
In all experiments, we use stratified 5-fold cross-validation,
to maintain the imbalance ratio across each folds. 
%%% Reviewers comment : how the imbalance in data is addressed
The imbalance ratio for both arousal and valence is quite high with a distribution of ($A_{high}$=818, $A_{low}$=468) and ($V_{high}$=808, $V_{low}$=472). Inorder to overcome the biased learning, we incorporated the class weights using python's $sklearn$ utility. 
%which automatically adjusts weights inversely proportional to class frequencies, into the cost function of the model.
%Table \ref{tab:res2} presents the weighted performance measures on different feature sets using feature selection as discussed in section \ref{FS}.
Table \ref{table:FS} represents the features selected from each of four sets followed by bands (/IMFs) selection, and reduced feature dimension. For MHS and HHSA, we considered first 4 IMFs for both arousal and valence, $n\_bins$ are selected as discussed in section \ref{FE}, followed by flattening the output 2D-matrix to feed the classifiers used. Thus, MHS and HHSA resulted into vector of dimension 4*64*32=8192-D and 5*5*32=800-D. We further reduced the dimension of MHS by taking only 10 frontal channels \cite{mohammadi2017wavelet}, giving the vector of 2560-D.

%%%% Reviewer comment : description of the parameters tuned
The hyperparameter of classifiers (KNN, SVM, RF) are tuned using $Grid Search$, where we found KNN with $K$=$3$ to be best among $K={3,5,8}$. For SVM, $rbf$ kernel performed best. 
We tuned RF with grid of parameters $estimators$ and $depth$ with a set of values where former varying from 50 to 300 with an increment of 50 and latter varying from 6 to 18 with an increment of 2 w.r.t each feature set.
From the results presented in Table \ref{tab:res2}, we found the PSDs with RF i.e RF ($estimators$=120, $estimators$=8) and set\_A features with RF ($estimators$=210, $estimators$=8) to be performing best in terms of weighted F1-score of 67.02\% and 65.73\%; Classification Accuracy (CA) of 67.42\% and 66.56\%, on valence and arousal, respectively.

It is to be noted that DWT and EMD-based features; HHSA specifically, trained on RF, performs competitively as compared to the best ones.
Hence, we perform late fusion on the posteriors of best performing models,  trained individually on each feature set. 
As shown in Table \ref{tab:res2}, RF performs best for all the feature sets on both arousal and valence, except $set\_A$ in valence for which SVM worked best.
Fusing posteriors further improved the performance, with weighted F1-score of 67.02\% and 66.6\%; CA of 68.8\% and 67.5\% (as shown in Table \ref{tab:sota}), on valence and arousal, respectively. 
%This improved the performance further, giving the best weighted F1-score and Classification Accuracy (CA) of  67.2\% and 68.8\% on valence and 66.6\% and 67.5\% on arousal, respectively. 

Due to varied experimental setting in terms of evaluation method, classification task, dataset etc., we couldn't directly compare the earlier EMD based work in EEG Emotion Recognition \cite{ji2019eeg,zhang2016approach}.
In Table \ref{tab:sota}, we compare our approach with existing state-of-the-art work on Emotion Recognition using DEAP data with similar settings as ours i.e subject-dependant and binary class classification. 
%For fair comparison, we report the results with 10-fold cross validation as well.
Our approach performed best among all for the valence. For arousal, while \cite{chao2019emotion} performs better by 0.8\%, it is important to be noted that segments used in train and test are from same trials which degrades drastically if tested on unseen trials. Whereas our evaluation method (like \cite{li2015eeg,huang2017eeg}) is trial-independant.

\vspace{-0.4cm}
\section{Conclusion}
\vspace{-0.2cm}
\label{conclusion}
%To be updated
In this paper, we address trial-independant EEG-based emotion recognition in A-V space. We explored the richness of EMD by analysing MHS and HHSA.
Further, we extract the multi-domain feature extraction based on DWT, EMD along with several other temporal and spectral features. 
Both DWT and EMD being complementary, result in boosting the performance. To investigate the effectiveness of these features with several other temporal and spectral features for emotion classification using EEG, an extensive experiment has been performed.
Experiments conducted on DEAP and comparison made with earlier work indicates the efficacy of our proposed approach. 
%HHSA is one of the most recent technique developed over EMD that give full informational representation of nonlinear signal. 
To best of our knowledge, this is the first attempt to use HHSA for recognizing emotion in EEG.
In future efforts, tempo-spatial deep models shall be investigated to capture the 2-D HHSA representation more efficiently.

%\newpage
\bibliographystyle{IEEEbib}
\bibliography{mybib}

%\bibliographystyle{IEEEbib}
%\bibliography{mybib}

\end{document}